\documentclass[floatfix,10pt,twocolumn,superscriptaddress,aps,prd,nofootinbib]{revtex4-2}
\usepackage{graphicx}
\usepackage{subfigure}
\usepackage{dcolumn}\usepackage{bm}
\usepackage[usenames]{color}
\usepackage{lineno}
\usepackage{amsmath}
\usepackage{amssymb}
\usepackage{etoolbox} 
\usepackage{url} 
\usepackage{placeins}
\usepackage{hyperref}

\usepackage{multirow}

\newcommand*\linenomathpatch[1]{%
  \cspreto{#1}{\linenomath}%
  \cspreto{#1*}{\linenomath}%
  \csappto{end#1}{\endlinenomath}%
  \csappto{end#1*}{\endlinenomath}%
}
\linenomathpatch{equation}
\linenomathpatch{gather}
\linenomathpatch{multline}
\linenomathpatch{align}
\linenomathpatch{alignat}
\linenomathpatch{flalign}

\def\address{\@ifstar{\address@star}%
  {\@ifnextchar[{\address@optarg}{\address@noptarg}}}

\newcommand\ee{e^+e^-}

\newcommand\gbl{g_{B-L}}

\newcommand\g{\gamma}
\newcommand\ma{m_{Z'}}

\newcommand\Na{{N}_{Z'}}

\newcommand\emu{e^- Z \to e^- Z \g; \g \to \mu^+ \mu^-}
\newcommand\mm{\mu^+ \mu^-}
\newcommand\ez{e^- Z \to e^- Z Z'}
\newcommand\eznu{e^- Z \to e^- Z Z';~ Z' \to \nu \nu}

\newcommand\nn{\nu \nu}
\newcommand\ec{{\rm ECAL}}
\newcommand\hc{{\rm HCAL}}
\newcommand\nz{n_{Z'}}
\newcommand\mz{m_{Z'}}

\newcommand\vt{{\rm VETO}}
\newcommand\ps{{\rm PS}}

\begin{document}

\title{
EUROPEAN LABORATORY FOR PARTICLE PHYSICS\\
\vskip0.5cm
\hspace{-3.0cm}{\rightline{\rm  CERN-EP-2022-156}}
\vskip1.cm
Search for a New $B-L$  $Z'$ Gauge Boson  with the NA64 Experiment at CERN \\
}
\author{Yu.~M.~Andreev}\affiliation{ Institute for Nuclear Research, 117312 Moscow, Russia}
\author{D.~Banerjee}\affiliation{ CERN, European Organization for Nuclear Research, CH-1211 Geneva, Switzerland}
\author{B.~Banto Oberhauser}\affiliation{ ETH Z\"urich, Institute for Particle Physics and Astrophysics, CH-8093 Z\"urich, Switzerland}
\author{J.~Bernhard}\affiliation{ CERN, European Organization for Nuclear Research, CH-1211 Geneva, Switzerland}
\author{P.~Bisio}\affiliation{INFN, Sezione di Genova, 16147 Genova, Italia}\affiliation{Universit\'a degli Studi di Genova, 16126 Genova, Italy}
\author{V.~E.~Burtsev}\affiliation{ Joint Institute for Nuclear Research, 141980 Dubna, Russia}
\author{A.~Celentano}\affiliation{INFN, Sezione di Genova, 16147 Genova, Italia}
\author{N.~Charitonidis}\affiliation{ CERN, European Organization for Nuclear Research, CH-1211 Geneva, Switzerland}
\author{A.~G.~Chumakov}\affiliation{ Tomsk Polytechnic University, 634050 Tomsk, Russia}
\author{D.~Cooke}\affiliation{ UCL Departement of Physics and Astronomy, University College London, Gower St. London WC1E 6BT, United Kingdom}
\author{P.~Crivelli}\affiliation{ ETH Z\"urich, Institute for Particle Physics and Astrophysics, CH-8093 Z\"urich, Switzerland}
\author{E.~Depero}\affiliation{ ETH Z\"urich, Institute for Particle Physics and Astrophysics, CH-8093 Z\"urich, Switzerland}
\author{A.~V.~Dermenev}\affiliation{ Institute for Nuclear Research, 117312 Moscow, Russia}
\author{S.~V.~Donskov}\affiliation{ State Scientific Center of the Russian Federation Institute for High Energy Physics of National Research Center 'Kurchatov Institute' (IHEP), 142281 Protvino, Russia}
\author{R.~R.~Dusaev}\affiliation{ Tomsk Polytechnic University, 634050 Tomsk, Russia}
\author{T.~Enik}\affiliation{  Joint Institute for Nuclear Research, 141980 Dubna, Russia}
\author{V.~N.~Frolov}\affiliation{  Joint Institute for Nuclear Research, 141980 Dubna, Russia}
\author{A.~Gardikiotis}\affiliation{ Physics Department, University of Patras, 265 04 Patras, Greece}
\author{S.~G.~Gerassimov }\affiliation{ Technische Universit\"at M\"unchen, Physik  Department, 85748 Garching, Germany}\affiliation{ P.N.Lebedev Physical Institute of the Russian Academy of Sciences, 119 991 Moscow, Russia}
\author{S.~N.~Gninenko}\thanks{Corresponding author}\email{Sergei.Gninenko@cern.ch}\affiliation{ Institute for Nuclear Research, 117312 Moscow, Russia}
\author{M.~H\"osgen}\affiliation{ Universit\"at Bonn, Helmholtz-Institut f\"ur Strahlen-und Kernphysik, 53115 Bonn, Germany}
\author{M.~Jeckel}\affiliation{ CERN, European Organization for Nuclear Research, CH-1211 Geneva, Switzerland}
\author{V.~A.~Kachanov}\affiliation{ State Scientific Center of the Russian Federation Institute for High Energy Physics of National Research Center 'Kurchatov Institute' (IHEP), 142281 Protvino, Russia}
\author{A.~E.~Karneyeu}\affiliation{ Institute for Nuclear Research, 117312 Moscow, Russia}
\author{G.~Kekelidze}\affiliation{  Joint Institute for Nuclear Research, 141980 Dubna, Russia}
\author{B.~Ketzer}\affiliation{ Universit\"at Bonn, Helmholtz-Institut f\"ur Strahlen-und Kernphysik, 53115 Bonn, Germany}
\author{D.~V.~Kirpichnikov}\affiliation{ Institute for Nuclear Research, 117312 Moscow, Russia}
\author{M.~M.~Kirsanov}\affiliation{ Institute for Nuclear Research, 117312 Moscow, Russia}
\author{V.~N.~Kolosov}\affiliation{ State Scientific Center of the Russian Federation Institute for High Energy Physics of National Research Center 'Kurchatov Institute' (IHEP), 142281 Protvino, Russia}
\author{S.~G.~Kovalenko}\affiliation{Center for Theoretical and Experimental Particle Physics, Facultad de Ciencias Exactas,
Universidad Andres Bello, Fernandez Concha 700, Santiago, Chile }\affiliation{Millennium Institute for Subatomic Physics at High-Energy Frontier (SAPHIR), Fernandez Concha 700, Santiago, Chile}
\author{V.~A.~Kramarenko}\affiliation{  Joint Institute for Nuclear Research, 141980 Dubna, Russia}\affiliation{ Skobeltsyn Institute of Nuclear Physics, Lomonosov Moscow State University, 119991  Moscow, Russia}
\author{L.~V.~Kravchuk}\affiliation{ Institute for Nuclear Research, 117312 Moscow, Russia}
\author{ N.~V.~Krasnikov}\affiliation{  Joint Institute for Nuclear Research, 141980 Dubna, Russia}\affiliation{ Institute for Nuclear Research, 117312 Moscow, Russia}
\author{S.~V.~Kuleshov}\affiliation{Center for Theoretical and Experimental Particle Physics, Facultad de Ciencias Exactas,
Universidad Andres Bello, Fernandez Concha 700, Santiago, Chile }\affiliation{Millennium Institute for Subatomic Physics at High-Energy Frontier (SAPHIR), Fernandez Concha 700, Santiago, Chile}
\author{V.~E.~Lyubovitskij}\affiliation{ Tomsk Polytechnic University, 634050 Tomsk, Russia}\affiliation{Millennium Institute for Subatomic Physics at High-Energy Frontier (SAPHIR), Fernandez Concha 700, Santiago, Chile}
\author{V.~Lysan}\affiliation{  Joint Institute for Nuclear Research, 141980 Dubna, Russia} 
\author{L.~Marsicano}\affiliation{INFN, Sezione di Genova, 16147 Genova, Italia} 
\author{V.~A.~Matveev}\affiliation{  Joint Institute for Nuclear Research, 141980 Dubna, Russia}
\author{Yu.~V.~Mikhailov}\affiliation{ State Scientific Center of the Russian Federation Institute for High Energy Physics of National Research Center 'Kurchatov Institute' (IHEP), 142281 Protvino, Russia}
\author{L.~Molina Bueno}\affiliation{ ETH Z\"urich, Institute for Particle Physics and Astrophysics, CH-8093 Z\"urich, Switzerland}\affiliation{Instituto de Fisica Corpuscular (CSIC/UV), Carrer del Catedrátic José Beltrán Martinez, 2, 46980 Paterna, Valencia}
\author{D.~V.~Peshekhonov}\affiliation{  Joint Institute for Nuclear Research, 141980 Dubna, Russia}
\author{V.~A.~Polyakov}\affiliation{ State Scientific Center of the Russian Federation Institute for High Energy Physics of National Research Center 'Kurchatov Institute' (IHEP), 142281 Protvino, Russia}
\author{B.~Radics}\affiliation{ ETH Z\"urich, Institute for Particle Physics and Astrophysics, CH-8093 Z\"urich, Switzerland}
\author{A.~Rubbia}\affiliation{ ETH Z\"urich, Institute for Particle Physics and Astrophysics, CH-8093 Z\"urich, Switzerland}
\author{K.~M.~Salamatin}\affiliation{ Joint Institute for Nuclear Research, 141980 Dubna, Russia}
\author{V.~D.~Samoylenko}\affiliation{ State Scientific Center of the Russian Federation Institute for High Energy Physics of National Research Center 'Kurchatov Institute' (IHEP), 142281 Protvino, Russia}
\author{H.~Sieber}\affiliation{ ETH Z\"urich, Institute for Particle Physics and Astrophysics, CH-8093 Z\"urich, Switzerland}
\author{D.~Shchukin}\affiliation{ P.N.Lebedev Physical Institute of the Russian Academy of Sciences, 119 991 Moscow, Russia}
\author{O.~Soto}\affiliation{Departamento de F\'{i}sica, Facultad de Ciencias, Universidad de La Serena, Avenida Cisternas 1200, La Serena, Chile}\affiliation{Millennium Institute for Subatomic Physics at High-Energy Frontier (SAPHIR), Fernandez Concha 700, Santiago, Chile}
\author{V.~O.~Tikhomirov}\affiliation{ P.N.Lebedev Physical Institute of the Russian Academy of Sciences, 119 991 Moscow, Russia}
\author{I.~V.~Tlisova}\affiliation{ Institute for Nuclear Research, 117312 Moscow, Russia} 
\author{A.~N.~Toropin}\affiliation{ Institute for Nuclear Research, 117312 Moscow, Russia}
\author{B.~I.~Vasilishin}\affiliation{ Tomsk Polytechnic  University, 634050 Tomsk, Russia}
\author{P.~V.~Volkov}\affiliation{  Joint Institute for Nuclear Research, 141980 Dubna, Russia}\affiliation{ Skobeltsyn Institute of Nuclear Physics, Lomonosov Moscow State University, 119991  Moscow, Russia}
\author{V.~Yu.~Volkov}\affiliation{ Skobeltsyn Institute of Nuclear Physics, Lomonosov Moscow State University, 119991  Moscow, Russia}
\author{I.~Voronchikhin}\affiliation{ Tomsk Polytechnic University, 634050 Tomsk, Russia}
\author{J. Zamora-Sa\'{a}}\affiliation{Center for Theoretical and Experimental Particle Physics, Facultad de Ciencias Exactas,
Universidad Andres Bello, Fernandez Concha 700, Santiago, Chile }\affiliation{Millennium Institute for Subatomic Physics at High-Energy Frontier (SAPHIR), Fernandez Concha 700, Santiago, Chile}

%
\collaboration{The NA64 Collaboration}\noaffiliation
\vskip 0.25cm

\date{\today}

\begin{abstract}  
A  search for a new $Z'$ gauge boson associated with (un)broken $B-L$ symmetry in the keV-GeV mass range is carried out for the first time  using the  missing-energy technique
in the NA64  experiment  at the CERN SPS. From the analysis of the data  with $3.22\times10^{11}$ electrons on target collected during 2016 - 2021 runs
  no signal events were found. This  allows to derive new constraints on the $Z'-e$ coupling strength, which for the mass range  $0.3\lesssim m_{Z'}\lesssim 100$ MeV are more stringent  compared to those  obtained from the neutrino-electron scattering data. 
\end{abstract}


\maketitle
Models with the gauged difference between baryon and lepton number, $B-L$, are attractive and well-motivated extensions of the standard model (SM) 
\cite{dav,mm}  that may explain two of the most challenging problems in particle physics today - the origin of neutrino masses \cite{mink} -\cite{glash} 
and the nature of dark matter (DM) \cite{kan}-\cite{mo}. They could also serve as an explanation for several existing experimental anomalies, such as e.g. an
excess of low energy events recently observed by XENON1T \cite{bo,lin1,choi}.
Among the possible realizations of such models, the minimal one is based on the gauge group $SU(2)_L \times U(1)_Y \times U(1)_{B-L}$, which simply 
extends the SM with an extra $U(1)$ gauge group associated to the difference of baryon number $B$ and lepton number $L$. In these models, the cancellation of gauge anomalies is usually achieved by adding three right-handed neutrinos, which simultaneously allow to explain neutrino masses via the type-I seesaw mechanism.  Below, we consider two  cases of 
$B-L$ extensions of the SM, with unbroken and spontaneously broken $B-L$ symmetry. The $B-L$ gauge coupling $g_{B-L}$  in these models can be small and  the associated gauge boson $Z'$ could have a mass well below the  electroweak scale ($\ll 100$ GeV), see e.g. \cite{heeck1,cam,oka3}.
 Searches for new physics at such a low energy scale recently received significant  attention from the community, see e.g. \cite{batta,pbc-bsm,agra}.
\par  In addition to Dirac neutrinos, an unbroken $U(1)_{B-L}$  brings with it only one more particle: the gauge boson $Z'$, coupled to the $B-L$ current 
$j_{B-L}= j_B - j_L$ via $g_{B-L} Z'_\mu  j^\mu_{B-L}$, and  leading to 
 the Lagrangian:
\begin{equation}
\mathcal{L} \supset  \gbl Z'_\mu \sum_{families} \bigl[ \frac{1}{3} \overline{q} \gamma^\mu q  - \overline{l} \gamma^\mu l - 
\overline{\nu} \gamma^\mu \nu \bigr]
\label{eq:lagr}
\end{equation}
where the $\gbl$ is the $U(1)_{B-L}$ coupling constant, and $q,~l$ and $\nu$ are quark, charged lepton and neutrino fields, respectively.
The $Z'$ can kinetically mix with the hypercharge boson \cite{holdom}, effectively coupling it to the hypercharge current. 
We will neglect this kinetic mixing in the following for simplicity.
\par For comparison with the  unbroken $B-L$, we will also consider the case of the spontaneously broken $B-L$ symmetry.
The neutrino sector of such gauge $B-L$ model consists of three heavy ($N_i$)  and three light Majorana neutrinos. The $N_i$ could be a viable 
 DM candidate explaining the relic density via the freeze-out mechanism with a mass lying in the range including the keV to TeV scale, see, e.g. \cite{hslee}.  
 \par   If the light $Z'$ boson exists, crucial questions about its mass scale, coupling constants, decay modes, etc. arise, providing  an important target for 
the ($m_{Z'};g_{B-L}$)  parameter space,  which can be probed at energies attainable at accelerators. One possible way to answer these questions is to
search for  $Z'$ in neutrino-electron scattering  experiments. The $Z'$ signature would be an observation of an excess of recoil electrons in  neutrino-electron scattering due
 to nonstandard $\nu-e$ interaction
 transmitted by the $Z'$. The signal event rate in the detector in this case scales as $\sim \frac{g_{B-L}^4 Z m_e E_\nu}{m_{Z'}^4}$, where $g_{B-L},~m_e,~E_\nu $ are the $Z'$ coupling strength, the mass of electron, and the neutrino energy, respectively,  and $Z$ is the charge of the target nuclei.
 Recently, severe limits on the $B-L~Z'$ excluding the  coupling strength range $ 10^{-6} \lesssim g_{B-L} \lesssim  10^{-2}$ 
 have been obtained for the masses  $1~{\rm keV} \lesssim m_{Z'} \lesssim 1$ GeV  \cite{aliev, lin1, lin2} from the results of neutrino-electron scattering experiments, such as 
TEXONO \cite{tex-csi,tex-hpge,tex-npcge} and GEMMA \cite{gemma} at  nuclear reactors, BOREXINO \cite{borexino} using solar neutrinos, LSND \cite{lsnd}, 
and CHARM II \cite{charm} at neutrino beams from accelerators, leaving,  however,  a significant area of the parameter space still unexplored. 
\par Another approach, considered in this work was  proposed in Refs.~\cite{Gninenko:2013rka,Andreas:2013lya,gkkk,gkkketl}. It  
 is based on the searches for invisible $Z'$ in missing energy events from the reaction chain $\eznu$ of the  bremsstrahlung $Z'$
 production in high-energy electron scattering off heavy nuclei of an active beam-dump and its subsequent prompt invisible decay into a neutrino pair.
 The advantage of this type of experiment compared to the
 neutrino-scattering is that its sensitivity is proportional to the square of the ratio of the $Z'$ coupling strength to its mass, 
$(\frac{g_{B-L}Z}{m_{Z'}})^2$,  associated with the  $Z'$ production in the primary reaction. In the former case, for $m_{Z'} \gg E_\nu$ and couplings $g_{B-L} \ll 1$, it is significantly suppressed by 
the additional factor $(g_{B-L}/m_{Z'})^2$, associated with the $Z'$ mediating the $\nu-e$ interaction. 
 \par   As the $Z'$ boson couples to any  fermion $f_i$ including the heavy neutrino $N_i$, the total decay width of $Z'$ is defined by the sum 
$\Gamma_{tot}(Z') =   \sum_{i}  \Gamma_i (Z'\to f_i \overline{f}_i)$
 over the leptonic invisible $\Gamma(Z' \to \nu_i\nu_i, N_i,N_i)$ and  visible $\Gamma(Z'\to \ee,~ \mm)$,  and hadronic 
$\Gamma(Z'\to q_i \overline{q}_i)$  final states.  The decay rate into a leptonic pair is given by
\begin{equation}
\Gamma_l(Z') = \frac{1}{3} \alpha_{B-L} m_{Z'} \bigl(1+2\frac{m_l^2}{m_{Z'}^2} \bigr) \bigl(1-4\frac{m_l^2}{m_{Z'}^2} \bigr)^{1/2} 
\label{eq:lepton}
\end{equation}
Here, $\alpha_{B-L}=\frac{g^2_{B-L}}{4\pi}$.
For the model with unbroken $U(1)_{B-L}$, the invisible width of  $Z'$ is then determined by its decay  into the three light Dirac neutrinos 
$\nu= \nu_L+\nu_R$,  $\Gamma_{inv} (Z') = 3 \Gamma(Z' \to \overline{\nu} \nu) = \alpha_{B-L} m_{Z'}$,
which effectively counts the number of light neutrinos. For the broken $U(1)_{B-L}$ case, in addition to the $Z'\to \nu \nu$ decays to light Majorana neutrinos, 
   three  invisible decays  of $Z'$ to  heavy Majorana neutrinos contribute assuming  the  $m_{Z'} > 2 m_{N_i}$ case with the rate  
\begin{equation}
\Gamma_{inv} (Z' \to N_i N_i) = \frac{1}{6} \alpha_{B-L} m_{Z'}  \bigl(1-4\frac{m_{N_i}^2}{m_{Z'}^2} \bigr)^{3/2} 
\label{eq:nuh}
\end{equation}
 \begin{figure}[tbh!!]
 \centering
\includegraphics[width=.45\textwidth]{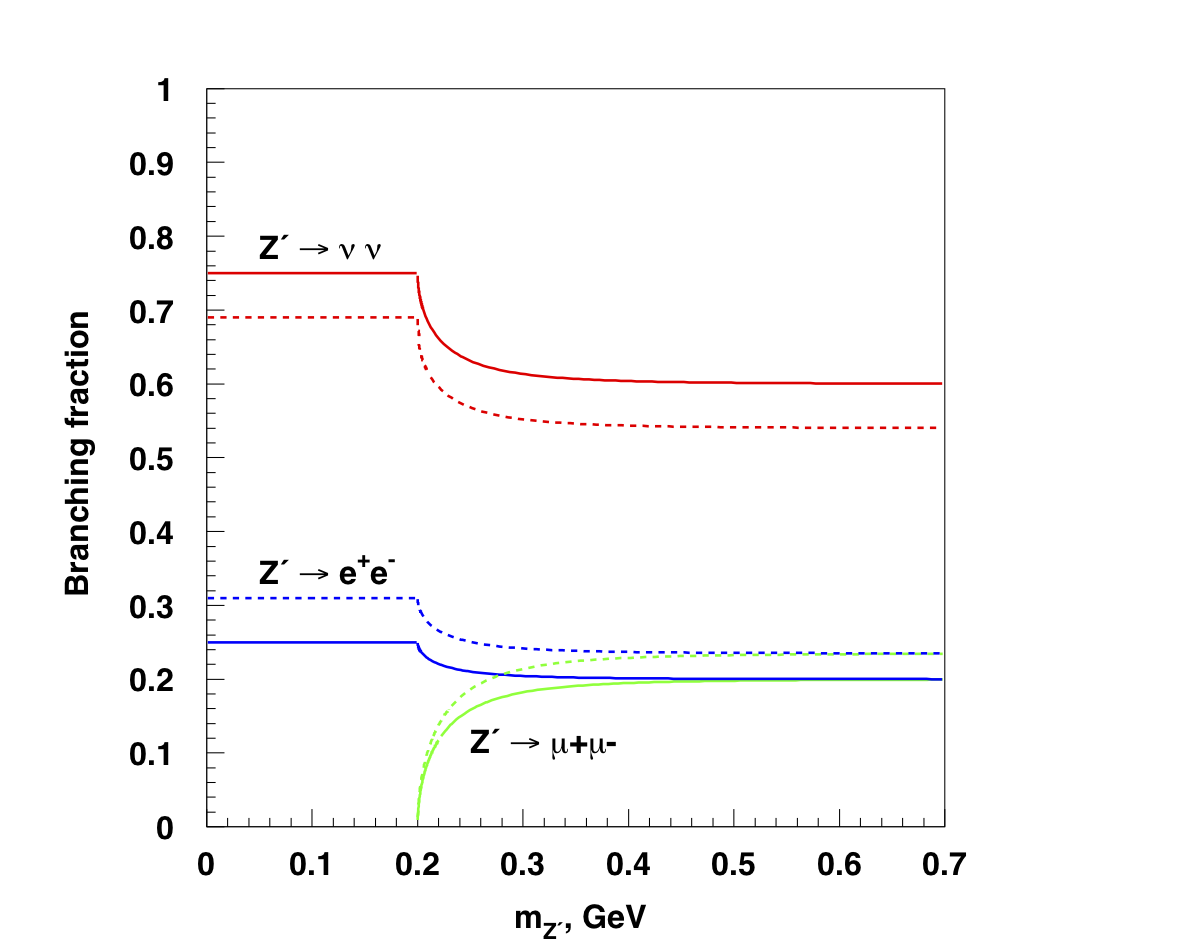}%
\caption{$Z'$ decay branching ratios to $\nu_i \nu_i; N_i N_i$(red), $\ee$(blue), and $\mm$ (green) for the unbroken (solid line) and broken (dashed line) $U(1)_{B-L}$ models. See text for details. \label{fig:bratio}}
\end{figure} 
 In the mass range  $m_{Z'} < 0.7$ GeV relevant for this work,  the $Z'$ decays  mostly
 invisibly  with the contributions from the visible decay modes  $Z'\to \ee, \mm$ the partial widths of which are shown in Fig.~\ref{fig:bratio}.
Here for simplicity,  three degenerate heavy neutrino species  with  the mass ratio $\frac{m_{N_i}}{m_{Z'}} =\frac {1}{3}$ used for calculation of the phase space factor in \eqref{eq:nuh} have been assumed.
  \par In this Letter we report new results on the search for  the $Z'$ in the NA64 fixed-target  experiment  at the CERN SPS.
The experiment  employed  the optimized H4  100 GeV  electron beam with a maximal intensity $\simeq 10^7$ electrons per  SPS spill of 4.8 s produced by the primary 400 GeV proton beam. The detector used 
the beam  scintillator and veto counters, a magnetic spectrometer  consisting of two successive  dipole magnets MBPL  and a low-material-budget tracker. The tracker was a set of micromegas (MM), straw-tube (ST) and GEM chambers  allowing the measurements of $e^-$ momenta with the precision $\delta p/p \simeq 1\%$ \cite{Banerjee:2015eno,st}. 
 Synchrotron radiation (SR) emitted in the MBPL magnetic field  
 was used for  efficient  tagging of beam electrons with a SR detector (SRD) \cite{Gninenko:2013rka, na64srd}, providing powerful suppression of the initial hadron contamination in the  beam $\pi/e^- \lesssim 10^{-2}$  down to the level  $\simeq 10^{-5}$.   
The detector was also equipped with an active target, which was an  electromagnetic calorimeter (ECAL),  a  matrix of  Shashlik-type modules  for  measurement of the electron energy  $E_{ECAL}$.  Each module has $\simeq 40$ radiation  lengths ($X_0$) with the first 4$X_0$  used as a preshower detector.   
 Downstream of the ECAL, the detector was equipped with a large  veto counter VETO, and a hadronic calorimeter (HCAL) of $\simeq 30$ nuclear interaction lengths. The HCAL served as an efficient veto to detect muons and  hadronic secondaries produced in the $e^- A$ interactions  in the target.  
 The events were collected with a beam defining  trigger  requiring, also,  an in-time cluster in the ECAL with the energy $E_{ECAL} \lesssim 80$ GeV. 
 More detail of the NA64 detector can be found in \cite{na64prl17,na64prd18,na64prl19}.
\par  The search described in this paper uses the data samples of $n_{EOT}=3.22\times 10^{11}$  electrons on target (EOT),  collected  in  the years 2016, 2017, 2018 \cite{na64prl17,na64prd18,na64prl19,na64alp} and 2021  with the beam intensities  mostly  in the range $\simeq (5-6)\times 10^6$    e$^-$ per spill.
 Data  from these four runs (hereafter called respectively  runs I,II, III, and IV) were processed with  selection criteria similar to the one used in 
 Refs. \cite{na64prd18, na64prl19} and, finally, combined as described below. Compared to the 2016-2018 runs, in the  2021 run the ECAL target and the HCAL were moved upstream to increase   the detector coverage resulting in a significant reduction of background from large-angle secondaries  from the $e^-$ hadronic interactions in the beam line. 
\par  A detailed  GEANT4 \cite{Agostinelli:2002hh, geant} based Monte Carlo (MC) simulation was used to study the signal acceptance and backgrounds, and  optimize selection criteria. For calculations of the signal yield we used the fully GEANT4 compatible package DMG4 
\cite{dmg4}. Using this package the production of $Z'$ in the process $\eznu$
  has been simulated with the cross sections obtained from the exact tree-level calculations, see, e.g., Refs. \cite{gkkk,gkkketl}. 
\par  The  number  $d \nz(\mz, E_{Z'}, E_0)$ of produced $Z'$ bosons with a given mass $m_{Z'}$ and energy $E_{Z'}$ per single EOT with the energy $E_0$  was obtained from
\begin{equation}
 \frac{d \nz}{dE_{Z'}} =  \frac{\rho N_{A}}{A_{{\rm Pb}}}  \int_0^{T}    \int_0^{E_0} n(E_0,E_e,s) \frac{d\sigma_{Z'}(E_{e})}{dE_{Z'}} dE_{e} ds
\label{z-number}
\end{equation} 
where  $ \rho$ is density of the target, $N_A$ is the 
Avogadro's number, $A_{{\rm Pb}}$ is the  
Pb atomic mass, $n(E_0,E_e,s)$ is  the number of $e^\pm$ in the {\it e-m} shower at the depth $s$ (in radiation lengths) with energy $E_e$  within the target of total thickness $T$, and 
$ \frac{d\sigma_{Z'}(E_{e})}{dE_{Z'}} $ is the differential cross section  for the $Z'$ production in the  reaction $\ez$ via the  interaction of Eq.(\ref{eq:lagr})  in the kinematically allowed region up to $Z'$ energies  $E_{Z'}\simeq E_e$ by an electron with the energy $E_e$. It  depends, in particular, on the coupling and mass  $\gbl, ~m_{Z'}$, and the beam energy $E_0$.
The efficiency of $\eznu$ event registration in our detector was also cross-checked by reconstructing the rare QED processes of dimuon production, $\emu$.
These events, dominated by the ECAL shower photons conversion into $\mm$ pairs on a target nucleus,  are similar to the $Z'\to \nn$ decay events if the 
energy deposition  in the HCAL is requested to be  above the two minimum ionizing particle threshold. The dimuon production
  was used  as a benchmark process allowing us to verify the reliability of simulations, systematic uncertainties and background estimations \cite{na64prd18, na64prl19}.  
  \par A blind analysis similar to the one described in 
Ref.\cite{na64prl19} was performed by using the following selection criteria: 
(i) The beam track momentum  should be within $100\pm 3$ GeV; 
(ii)  The energy detected by the SRD should be consistent with the SR energy emitted by $e^-$'s in the magnets;
(iii)  The shower shape  in the ECAL should be as expected from the signal-event  shower \cite{gkkk};
(iv) A single track should be reconstructed in the tracker chambers upstream of the ECAL; and (v) There should be no activity in the $\vt$.
\par
\begin{figure}[tbh!!]
 \centering
\includegraphics[width=.4\textwidth]{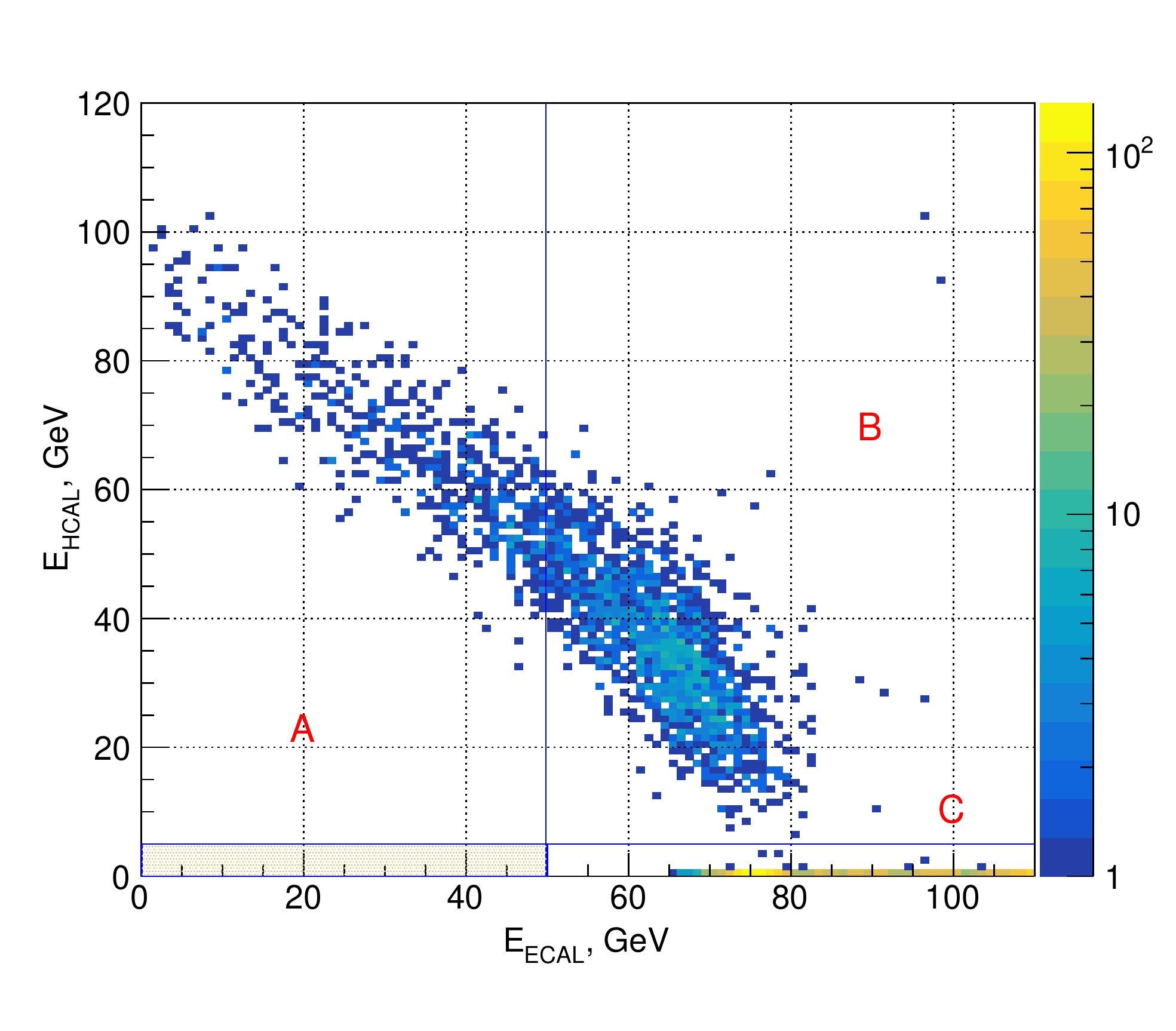}%
\caption{The  energy distribution of events in the $E_\ec;E_\hc$ plane selected from the data sample of the 2021 run
collected  with intensity $\simeq 5\times 10^6~e^-$ per spill  after applying all selection criteria. The shaded area is the signal box, which contains no events. The size of the signal box along the $E_\hc$ axis is increased by a factor of 5 for illustration.
\label{ecvshc}}
\end{figure}

 In the 2021 run the main background faking  the  signal of $\ez$ from the hadronic interactions of the $e^-$ beam  in the beam line materials 
accompanied by the emission of hadronic secondaries at a  large-angle (high $p_T$)  was more suppressed compared to the 2016-2018 runs  due to the improved  
detector coverage. By selecting events with no additional tracks or hits in MM and ST chambers upstream and  downstream of the magnets, most of events with charged hadronic secondaries were rejected.  The remaining background of $0.03\pm 0.015$ events from large angle secondary neutrals was evaluated directly from the data by the extrapolation of
events from the sideband ($E_{\ec} > 50~{\text GeV }; E_{\hc} < 1~{\text GeV }$) (region $C$ in Fig.\ref{ecvshc}) into the signal region and
estimating the systematic errors by varying the fit functions as described in Ref. \cite{na64prd18}.
Another background from  region $A$  in Fig.\ref{ecvshc}, mostly from  punch through of leading neutral hadrons $(n, K^0_L)$ with energy $\gtrsim 0.5~E_0$ produced by the beam $e^-$s  in the target, was evaluated from the study of their propagation through the HCAL modules \cite{na64alp} and was found to be negligible.
 Other sources of background such as loss of dimuons and decays in flight of beam $\pi$, $K$, were simulated and were also found
to be negligible. 
 \par  The overall signal efficiency $\epsilon_{Z'}$, which includes efficiencies for the  geometrical acceptance, the track, SRD, $\vt$ and HCAL
selections and the DAQ dead time, was found to be slightly $m_{Z'}$ dependent \cite{na64prl19}. The  ECAL signal selection efficiency, 
$\epsilon_{\ec}$, was estimated for different $Z'$ masses. The $\epsilon_{\ec}$ value for a shower from a $Z'$ event 
has to be corrected compared to the ordinary {\it e-m} shower, due to differences in the development of the {\it e-m} showers  at the early stage
in the ECAL preshower (PS) \cite{gkkk}. This correction was $\lesssim (5\pm 3)\%$, depending on the energy threshold in the PS ($E^{th}_{\ps}$) used in the trigger. 
The systematic uncertainty is dominated by the $E^{th}_{\ps}$ variation during the run, mostly due to the instabilities in the photomultilier gains.
The $\vt$ and HCAL selection efficiencies are defined by the noises, pile up and the leakage of the signal shower energy from the ECAL to these detectors.
They were studied using the electron calibration runs and simulations. 
The uncertainty in the efficiencies estimated to be $\lesssim 4\%$ is dominated mostly by the pile up effect.
The total signal selection efficiency with all criteria used except ECAL threshold on missing energy  varied from $\sim 0.62$ to $\sim 0.48$ with the uncertainty in the
signal yield to be $\simeq 10\%$ \cite{na64prd18}.
\par Data from runs I-IV were analyzed simultaneously using the multibin limit setting \cite{na64prd18} technique, with the code based on the
RooStats package \cite{root}. The signal box ($E_{\ec} < 50~{\text GeV }; E_{\hc} < 1~{\text GeV }$) was defined based on the energy spectrum calculations for $Z'$ bosons emitted by $e^\pm$ from the {\it e-m} shower generated by the primary  $e^-$s in the  ECAL \cite{gkkk, gkkketl}  and the HCAL zero-energy threshold determined mostly by the noise of the readout electronics. 
\begin{figure}[tbh!!]
\begin{center}
\includegraphics[width=0.5\textwidth]{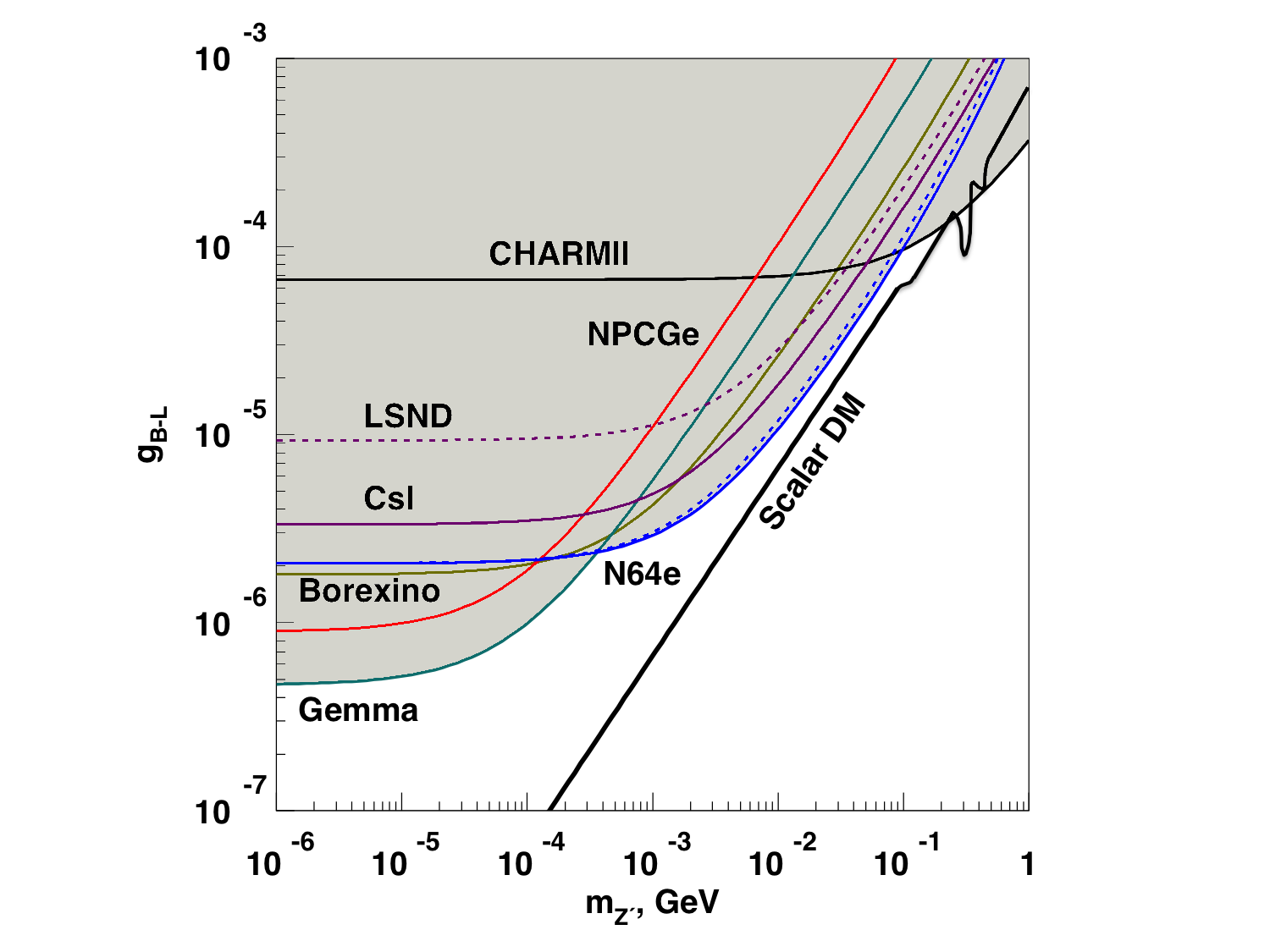}
\caption {The NA64 90\% C.L. exclusion region in the ($m_{Z'}, g_{B-L}$) 
plane for the unbroken (solid blue line) and broken (dashed blue line)  $U(1)_{B-L}$.  
For the latter case, three degenerate heavy neutrino species  with  the mass ratio $\frac{m_{N_i}}{m_{Z'}} = \frac{1}{3}$ have been assumed.
Constraints from the results of   neutrino-electron scattering experiments obtained with nuclear-reactor neutrinos at 
TEXONO \cite{tex-csi,tex-hpge,tex-npcge} and GEMMA \cite{gemma}, solar neutrinos at  BOREXINO \cite{borexino}, and accelerator  neutrino beams at 
LSND \cite{lsnd}, and CHARM II \cite{charm}  derived  in  Refs.  \cite{aliev, lin1} are also shown. The dashed area above the curves is excluded. As an example, the black curve 
illustrates  the parameter space for the thermal scalar DM  model for which the abundance of scalar $\chi$ is in agreement with the observed DM energy density. 
The curve is calculated assuming the $Z'-\chi$ coupling $\alpha_D=0.1$ and  the mass ratio $\mz = 3 m_\chi$,  see Ref.\cite{berlin}.   
\label{exclinv}}
\end{center}
\end{figure} 
The size of the signal box was optimized by comparing sensitivities defined as an average expected limit calculated using
the profile likelihood method. 
It was found weakly dependent on the $Z'$ mass and was finally set to $E_{\ec} \lesssim 50$ GeV for
all four runs and the whole mass range. The uncertainties in the signal yield and background level were treated as nuisance parameters in the statistical model \cite{Gross:2007zz}.
\par The total number of signal events in the signal box was the sum of events expected  from each of the four runs:
\begin{equation}
\Na= \sum_{j=1}^{4} n_{EOT}^j \epsilon_{Z'}^j n_{Z'}^j(g_{B-L},\ma, \Delta E_{Z'})
\label{eq:nev}
\end{equation}
where $\epsilon_{Z'}^j$ is the signal efficiency in run $j$, and $n_{Z'}^j(g_{B-L},\ma, \Delta E_{Z'})$ is the 
signal yield per EOT generated in the energy  range $\Delta E_{Z'} $.  Each $j$th entry in Eq.(\ref{eq:nev}) was calculated with simulations of signal events processing them through the reconstruction program  with the same selection criteria and  efficiency corrections as for the  data sample from run $j$.
The corresponding yield $n_{Z'}^j$ of $Z'\to invisible$  events was defined by 
\begin{eqnarray}
n_{Z'}^j (g_{B-L},\ma, \Delta E_{Z'})= \int \frac {dn_{Z'}^j}{dE_{Z'}} \bigl[Br(Z'\to \nn)+ \\ \nonumber
+\sum_{l=e,\mu}Br(Z'\to l^+l^-) e^{-\frac{ L_\ec +L_\hc}{L_{Z'}}} \bigr] dE_{Z'}  
\label{eq:events}
\end{eqnarray}
where $ \frac {dn_{Z'}^j}{dE_{Z'}}$ was calculated with Eq.(\ref{z-number}), and the  term in square brackets  gives the probability for the produced  $Z'$
 with a given mass $m_{Z'}$ and energy $E_{Z'}$, to make a transition  into the invisible final state, i.e. to decay either into
 $\nn$ pair, or outside the HCAL modules into a $\ee$ or $\mm$ pair. Here, $Br(Z'\to \nn)$  varies in the range 0.6 - 0.75, depending  on $m_{Z'}$, see Fig.~\ref{fig:bratio}, 
  $L_{Z'} = \frac{c\tau_{Z'} E_{Z'}}{m_{Z'}}$  and $ L_\ec +L_\hc$ are the $Z'$ decay length  and the total length of the ECAL + HCAL detectors, respectively, and 
   $\tau_{Z'}=\frac{1}{\Gamma_{tot}(Z')}$ is the lifetime of the $Z'$. After determining all selection criteria, background level, systematic uncertainties, we open the box and found 0 events, as shown in the left panel of Fig.~\ref{ecvshc}, consistent with $0.56\pm 0.17$ events from the background estimations for the full data sample
from the 2016-2018 \cite{na64prl19} and 2021 runs.  
\par The combined 90\% CL exclusion limits on the coupling  $\gbl$ as a function of the $Z'$ mass, calculated using the modified frequentist approach
\cite{na64prl19, junk,limit,Read:2002hq} for the models with unbroken and broken $U(1)_{B-L}$ symmetry are shown in Fig.~\ref{exclinv}. For the latter case, three degenerate heavy neutrino species  with  the mass ratio $\frac{m_{N_i}}{m_{Z'}} = \frac{1}{3}$ have been assumed. As an example, we also show the most motivated region of the parameter space
for the thermal scalar DM model, in which the $Z'$ also mediates new feeble interaction between the SM and DM \cite{berlin}. 
The treatment of the general case with other assumptions for the $N_i$  masses is straightforward and does not qualitatively change our main conclusions.
For example, if the $m_{Z'} < 2 m_{N_i}$ case is assumed, i.e. no $Z'$ decays into heavy neutrinos,  the  $Z'$  invisible branching ratio of $Br(Z'\to \nu_i\nu_i; N_i N_i) \simeq 54\%$ shown in Fig.~\ref{fig:bratio} drops to  $Br(Z'\to \nu_i\nu_i) \simeq 43\%$  and the corresponding limit for the mass range $m_{Z'} \gtrsim 100$ MeV, see Fig.~\ref{exclinv},  will be 
 $\simeq 10\%$ worse.  
Finally, note that  for the   mass range $0.3 \lesssim m_{Z'} \lesssim 100$ MeV, NA64 bounds  are more stringent than those derived  from  the results of neutrino-electron scattering. Another advantage of the NA64 approach compared to neutrino experiments in the case of the signal observation is its potential capability to distinguish the unbroken and broken $B-L$ scenarios  by measuring the ratio $\frac{Br(Z' \to invisible)}{Br(Z'\to \ee)}$, see Fig.~\ref{fig:bratio}.  
\par We gratefully acknowledge the support of the CERN management and staff  for their vital contributions. 
 This work was supported by the  Helmholtz-Institut f\"ur Strahlen-und Kernphysik (HISKP), University of Bonn (Germany), Joint Institute for Nuclear Research (JINR) (Dubna), the Ministry of Science and Higher Education (MSHE)  and RAS (Russia), ETH Zurich and SNSF Grants No. 169133, No. 186181, No. 186158,  No. 197346 (Switzerland), and FONDECYT Grants  No.1191103, No. 190845 and ANID$-$Millennium Program $-$ Grant No. ICN2019\_044 (Chile).  We acknowledge the support of the European Research Council (ERC) under the European Union's Horizon 2020 research and innovation programme, Grant No. 947715 (POKER).


\begin{thebibliography}{99}
  \bibitem{dav} A. Davidson, 
Phys. Rev. D {\bf 20}, 776 (1979).
\bibitem{mm} R.E. Marshak and R.N. Mohapatra, 
Phys. Lett.  {\bf 91B}, 222 (1980).
\bibitem{mink} P. Minkowski, 
Phys. Lett. {\bf 67B }, 421 (1977).
\bibitem{yan}T. Yanagida, 
Conf. Proc. C {\bf 7902131}, 95 (1979).
\bibitem{gell} M. Gell-Mann, P. Ramond, and R. Slansky, 
 Conf. Proc. C {\bf 790927}, 315 (1979).
\bibitem{ms} R.N. Mohapatra and G. Senjanovic, 
 Phys. Rev. Lett. {\bf 44}, 912 (1980).
\bibitem{glash} S.L. Glashow, 
 NATO Sci. Ser. B {\bf 61}, 687 (1980).
\bibitem{kan} S. Kanemura, T. Matsui,  and H. Sugiyama, 
 Phys. Rev. D {\bf 90},   013001 (2014).
\bibitem{oka1} N. Okada and O. Seto, 
Phys. Rev. D {\bf 82}, 023507 (2010).
\bibitem{oka2} N. Okada and Y. Orikasa,~
 Phys. Rev. D {\bf 85}, 115006 (2012).
\bibitem{19} L. Basso, O. Fischer, and J.J. van der Bij, 
Phys. Rev. D {\bf 87}, 035015 (2013).
\bibitem{20} T. Basak and T. Mondal, 
 Phys. Rev. D {\bf 89},  063527 (2014).
\bibitem{21} M. Lindner, D. Schmidt,  and A. Watanabe, 
Phys. Rev. D {\bf 89},  013007  (2014).
\bibitem{22} M. Duerr, P. Fileviez Perez,  and J. Smirnov, 
 Phys. Rev. D {\bf 92}, 083521 (2015).
\bibitem{23} J. Guo, Z. Kang, P. Ko,  and Y. Orikasa, 
Phys. Rev. D {\bf 91}, 115017  (2015).
\bibitem{24} W. Rodejohann and C.E. Yaguna, 
J. Cosmol. Astropart. Phys {\bf 12}, (2015) 032 .
\bibitem{mo} R.N. Mohapatra and N. Okada, 
Phys. Rev. D {\bf 102},  035028 (2020).
\bibitem{bo} C. Bohm, D. G. Cerdeno, M. Fairbairn, P. A. N. Machado, and A. C. Vincent, 
Phys. Rev. D {\bf 102}, 115013 (2020).
\bibitem{lin1} M. Lindner,  Y. Mambrini,  T. B. de Melo, and F. S. Queiroz, 
Phys. Lett. B {\bf 811},  135972 (2020).
\bibitem{choi} G. Choi, T. T. Yanagida, and N. Yokozaki, 
Phys. Lett. B {\bf 810}, 135836 (2020).
\bibitem{heeck1} J. Heeck, 
Phys. Lett. B {\bf 739}, 256 (2014).
\bibitem{cam}  M.D. Campos, D. Cogollo, M. Lindner, T. Melo, F. S. Queiroz, and W. Rodejohann, 
J. High Energy Phys. {\bf 08},  (2017) 092.
\bibitem{oka3} N. Okada, S. Okada, D. Raut, and Q.Shafi, 
Phys. Lett. B {\bf 810}, 135785 (2020). 
\bibitem{batta} M. Battaglieri {\it et al.},  arXiv:1707.04591.
\bibitem{pbc-bsm}
J.  Beacham  {\it et al.},  J.\ Phys.\ G {\bf 47},  010501 (2020).   
\bibitem{agra} P. Agrawal {\it et al.},  
 Eur. Phys. J. C {\bf 81}, 1015 (2021).
 \bibitem{holdom}
 B. Holdom, Phys. Lett.  {\bf 166B}, 196  (1986).
 \bibitem{hslee}
 K. Kaneta, Zh. Kang, and H.-S. Lee, JHEP {\bf 02}, 031  (2017).
\bibitem{aliev} S. Bilmis, I. Turan, T. M. Aliev, M. Deniz, L. Singh, and H. T. Wong, 
Phys. Rev. D {\bf 92}, 033009 (2015).
\bibitem{lin2} M. Lindner, F. S. Queiroz, W. Rodejohann, and Xun-Jie Xu, 
J. High Energy Phys. {\bf 05},  (2018) 098.
\bibitem{tex-csi}  M. Deniz et al., 
Phys. Rev. D {\bf 81}, 072001 (2010).
\bibitem{tex-hpge} H. B. Li {\it et al.}, 
Phys. Rev. Lett. {\bf 90}, 131802 (2003); 
Phys. Rev. D {\bf 75}, 012001 (2007).
\bibitem{tex-npcge} J.-W. Chen, H.-C. Chi, H.-B. Li, C.-P. Liu, L. Singh, H. T. Wong, C.-L. Wu, and C.-P. Wu, 
 Phys. Rev. D {\bf 90}, 011301(R) (2014).
\bibitem{gemma} A. G. Beda, E. V. Demidova, A. S. Starostin, V. B. Brudanin, V. G. Egorov, D. V. Medvedev, M. V. Shirchenko, and T. Vylov, 
Phys. Part. Nucl. Lett. {\bf 7}, 406 (2010).
\bibitem{borexino} G. Bellini {\it et al.}, 
Phys. Rev. Lett. {\bf 107}, 141302 (2011).
\bibitem{lsnd} L.B. Auerbach {\it et al.}, 
 Phys. Rev. D {\bf 63}, 112001 (2001).
\bibitem{charm} P. Vilain {\it et al.},
 Phys. Lett. B {\bf 302}, 351 (1993);  
 {\bf 335}, 246 (1994).
\bibitem{Gninenko:2013rka}
S.~N.~Gninenko,
  Phys.\ Rev.\ D {\bf 89},  075008 (2014).
 
\bibitem{Andreas:2013lya} 
  S.~Andreas {\it et al.},  arXiv:1312.3309.

\bibitem{gkkk} 
  S.~N.~Gninenko, N.~V.~Krasnikov, M.~M.~Kirsanov,  and D.~V.~Kirpichnikov,
  Phys.\ Rev.\ D {\bf 94},  095025 (2016).




  
 \bibitem{gkkketl}
  S.~N.~Gninenko, D.~V.~Kirpichnikov, M.~M.~Kirsanov,  and N.~V.~Krasnikov,   
   Phys. Lett. B {\bf 782}, 406 (2018).


\bibitem{Banerjee:2015eno}
  D.~Banerjee, P.~Crivelli,  and A.~Rubbia,
  Adv.\ High Energy Phys.\  {\bf 2015},  105730 (2015).

\bibitem{st} V.Yu. Volkov {\it et al.}, Phys. Part. Nucl. Lett. {\bf 16}, 847(2019).

\bibitem{na64srd}  
  E.~Depero {\it et al.},
  Nucl.\ Instrum.\ Methods. Phys. Res., Sect. A {\bf 866},  196 (2017).

 
 \bibitem{na64prl17} 
  D.~Banerjee {\it et al.} (NA64 Collaboration),
  Phys.\ Rev.\ Lett.\  {\bf 118},   011802 (2017).
 

\bibitem{na64prd18} 
  D.~Banerjee {\it et al.} (NA64 Collaboration),
  Phys.\ Rev.\ D {\bf 97},  072002 (2018).


 \bibitem{na64prl19} 
  D.~Banerjee {\it et al.} (NA64 Collaboration),
  Phys.\ Rev.\ Lett.\  {\bf 123},  121801 (2019).
  
\bibitem{na64alp} 
  D.~Banerjee {\it et al.} (NA64 Collaboration),
  Phys.\ Rev.\ Lett.\  {\bf 125},  081801 (2020).




  
  
\bibitem{Agostinelli:2002hh}
  S.~Agostinelli {\it et al.} [GEANT4 Collaboration],
 Instrum. Methods Phys. Res., Sect. {\bf 506}, 250  (2003).

\bibitem{geant} 
  J.~Allison {\it et al.},
  IEEE Trans.\ Nucl.\ Sci.\  {\bf 53}, 270 (2006).


\bibitem{dmg4}  M. Bondi, A. Celentano, R. R. Dusaev, D. V. Kirpichnikov, M. M. Kirsanov, N. V. Krasnikov, L. Marsicano, and D. Shchukin, 
Comput. Phys. Commun. {\bf 269}, 108129 (2021).
\bibitem{root} 
L. Moneta {\it et al.}, Proc. Sci.  ACAT2010, (2010) 057.






\bibitem{Gross:2007zz}
  E.~Gross,
 Report No. CERN-2008-001, CERN, 2008, p.71.
  
  
  \bibitem{junk}
T. Junk, 
Nucl. Instrum. Methods Phys. Res., Sect.{\bf 434}, 435 (1999).


\bibitem{limit}
G. Cowan, K. Cranmer, E. Gross, and O. Vitells, 
Eur. Phys. J. C {\bf 71}, 1554  (2011).

\bibitem{Read:2002hq}
  A.~L.~Read,
  J.\ Phys.\ G {\bf 28}, 2693 (2002).

 \bibitem{berlin}
 
  A. Berlin,  N. Blinov, G. Krnjaic,  P. Schuster, and N. Toro,  Phys.\ Rev.\ D {\bf 99},  075001 (2019).

\end{thebibliography}
\end{document}